\documentclass[sigconf]{acmart}

\copyrightyear{2021}
\acmYear{2021}
\setcopyright{acmcopyright}
\acmConference[SIGIR '21] {Proceedings of the 44th International ACM SIGIR Conference on Research and Development in Information Retrieval}{July 11--15, 2021}{Virtual Event, Canada.}
\acmBooktitle{Proceedings of the 44th International ACM SIGIR Conference on Research and Development in Information Retrieval (SIGIR '21), July 11--15, 2021, Virtual Event, Canada}
\acmPrice{15.00}
\acmISBN{978-1-4503-8037-9/21/07}
\acmDOI{10.1145/3404835.3462890}

\usepackage{multirow}
\usepackage{subfigure} 
\usepackage{lineno}
\usepackage[ruled]{algorithm2e}
\usepackage{color}
\usepackage{tabularx}
\usepackage{booktabs}

\newcolumntype{L}[1]{>{\raggedright\arraybackslash}p{#1}}
\newcolumntype{C}[1]{>{\centering\arraybackslash}p{#1}}
\newcolumntype{R}[1]{>{\raggedleft\arraybackslash}p{#1}}

\AtBeginDocument{%
  \providecommand\BibTeX{{%
    \normalfont B\kern-0.5em{\scshape i\kern-0.25em b}\kern-0.8em\TeX}}}

\begin{document}
\fancyhead{}

\title{StackRec: Efficient Training of Very Deep Sequential Recommender Models by Iterative Stacking}

\author{Jiachun Wang$^{1,4*}$, Fajie Yuan$^{2,3*}$, Jian Chen$^{1\dagger}$, Qingyao Wu$^{1\dagger}$, Min Yang$^{4\dagger}$, Yang Sun$^{4}$, \and Guoxiao Zhang$^{3}$}

\affiliation{\institution{$^{1}$School of Software Engineering, South China University of Technology}}
\affiliation{\institution{$^{2}$Westlake University \qquad $^{3}$Tencent}}

\affiliation{\institution{$^{4}$Shenzhen Institutes of Advanced Technology, Chinese Academy of Sciences}}

\email{sewangjiachun@mail.scut.edu.cn, yuanfajie@westlake.edu.cn, {ellachen, qyw}@scut.edu.cn,}

\email{{min.yang, yang.sun}@siat.ac.cn, suranzhang@tencent.com}

\thanks{$^{*}$Equal Contribution. This work was done when Jiachun Wang interned at SIAT, Chinese Academy of Sciences. This work was done when Fajie worked at Tencent (past affiliation) and Westlake University (current affiliation)
}
\thanks{$^{\dagger}$Corresponding authors.}


\begin{abstract}
Deep learning has brought great progress for the sequential recommendation (SR) tasks.
With advanced network architectures, sequential recommender models can  be stacked with many hidden  layers, e.g., up to 100 layers on real-world recommendation datasets.
Training such a deep network is difficult because it can be computationally very expensive and
takes much longer time, especially in situations where there are tens of billions of user-item interactions. To deal with such a challenge, we present StackRec, a simple, yet very effective and efficient training framework for deep SR models by iterative layer stacking. Specifically,
we first offer an important insight that hidden layers/blocks in a well-trained deep SR model have very similar distributions. Enlightened by this, we propose the \textit{stacking} operation on the pre-trained layers/blocks to transfer knowledge from a shallower model to a deep model, then we perform iterative stacking
 so as to yield a much deeper but easier-to-train SR model. We validate the performance of StackRec by instantiating it with four state-of-the-art SR models in three practical scenarios with  real-world datasets. Extensive experiments show that StackRec achieves not only comparable performance,
but also substantial acceleration in training time, compared to SR models that are trained from scratch. Codes are available at \textcolor{blue}{\url{https://github.com/wangjiachun0426/StackRec}}.

\end{abstract}

\ccsdesc[500]{Information systems~Recommender systems}
\ccsdesc[500]{Computing methodologies~Neural networks}

\keywords{Recommender systems; Knowledge Transfer; Training acceleration}

\maketitle

\begin{figure}[h!]
    \vspace{-0.1in}
	\centering
	\subfigure[40\% training data]{
		\includegraphics[width=0.47\linewidth]{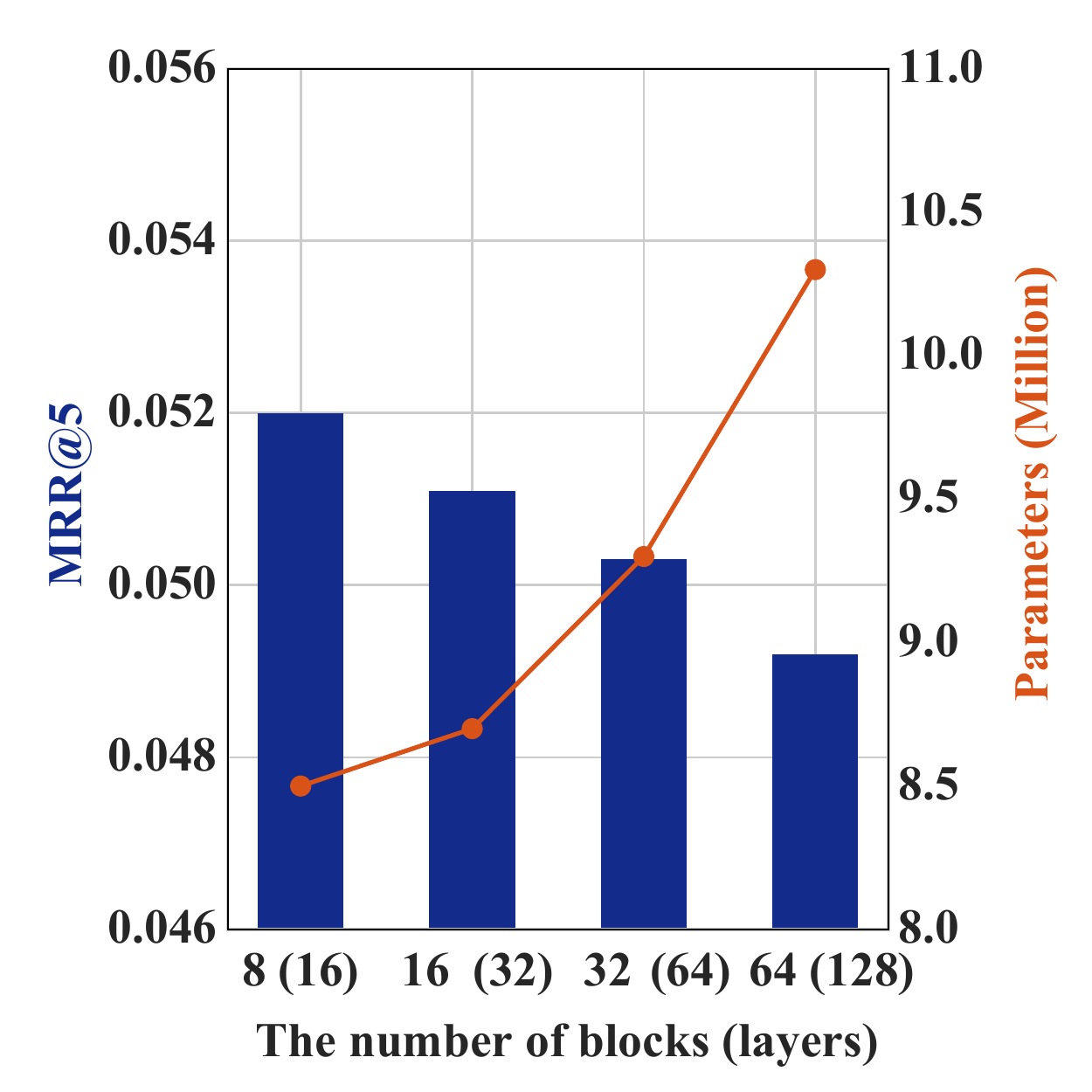}
	}
	\subfigure[100\% training data]{
		\centering
		\includegraphics[width=0.47\linewidth]{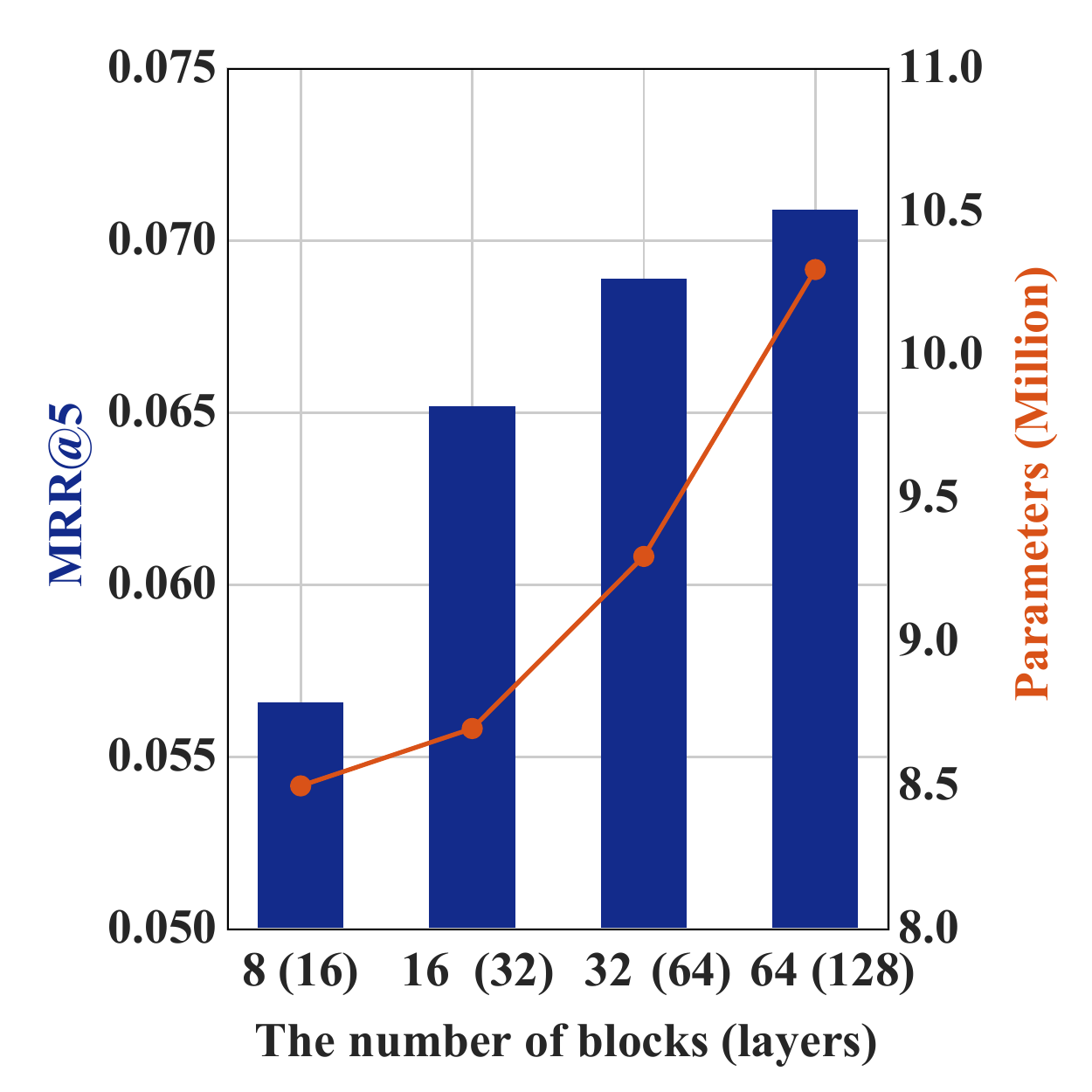}
	}
	\caption{Illustration of prediction accuracy  w.r.t. the model depth of NextItNet on Kuaibao (see Section~\ref{DataInfo} for details) using 40\% and 100\% of the original training data, respectively. }
	\label{increaselayers}
\end{figure}

\section{Introduction}
Recommender systems (RS) have
become a vital tool in  alleviating information overload, and have been
widely deployed for many online personalized applications. In particular, sequential (a.k.a., session-based) recommendations (SR) are becoming increasingly more popular since SR models are often more effective in modeling and inferring users' dynamic preference 
relative to traditional collaborative filtering (CF) methods. Moreover, SR models could also be trained by the self-supervised manner~\cite{yuan2019simple,kang2018self,yuan2020one,yuan2020parameter} --- i.e., predicting the next interaction in
the users' behavior sequence without human supervision of both features and labels
. Thanks to this, SR models could not only recommend items for the current system, but also generate 
high-quality user representations which are useful for solving the data scarcity problem of a different system or task~\cite{yuan2020parameter,yuan2020one}.


Despite the effectiveness, SR models require much deeper network architectures (than the standard CF) to fulfil their capacity in modeling complex relations of user behaviors and long-term user preference. As shown in Figure~\ref{increaselayers} (b),
the state-of-the-art SR model NextItNet~\cite{yuan2019simple} with a slight modification (see Section~\ref{nextitnet}) has to be stacked with around 128 convolutional layers for achieving its optimal performance. This observation actually surprises us as well since, except in~\cite{sun2020generic}, most
recommendation models in literature apply less than 10 layers for evaluation. On the other hand, by contrasting (a) with (b), we  can make another observation.
That is, the same SR model requires very different layer depths at different stages of the system ---  more intermediate (hidden) layers are often required for better accuracy when there are sufficient training examples.


Though a deeper \& larger network usually brings significant accuracy improvement, it may also result in expensive computational cost and longer training time even  using high-performance hardware, especially in real production systems where there are tens of billions of training data. 
In this paper, we aim to improve the training efficiency of very deep SR models in the \textbf{algorithmic} sense rather than assigning them more hardware resources. 
Our key motivation is from an interesting phenomena that intermediate layers/blocks of a deep SR model
have a highly similar distribution. The finding potentially suggests that knowledge (i.e., parameters)
of a shallow model could be reused by a deep model.
Specifically, suppose that we are given a well-optimized shallow model. Then we can apply the layer \textit{stack} operation by assigning weights of this shallow model to newly added top layers of the deep model. 
By fine-tuning, the double-depth SR model optimized in the stacking manner could achieve equivalent performance as the standard optimization method (i.e., training deep models from scratch). By iteratively stacking and fine-tuning, we could finally develop a very deep SR model.
Such stacked deep SR model substantially reduces
its training time
due to effective knowledge transfer.
We refer to the proposed efficient training framework for the SR task as StackRec.

StackRec can be applied into three  common recommendation scenarios.  
First, in a real production system, recommendation models may experience two periods: the data-hungry period (with limited users, items, and especially their interactions) at the initial stage of RS, and the data-sufficient period when the RS has accumulated over a long period of time~\cite{zhang2020retrain}. In the
former period, a shallow recommendation model is always desirable since deep models 
may suffer from serious overfitting and bring meaningless computation, as illustrated in Figure~\ref{increaselayers} (a). As new training data  comes in, the original shallow model might not be expressive enough any more, as illustrated in Figure~\ref{increaselayers} (b). As such, a new RS model with a deeper architecture should be rebuilt and optimized. In such a case, StackRec is of great help since it is able to transfer knowledge from the shallow model to an expected deep model rather than training it from scratch.
More ambitiously, real RS would eventually become lifelong learning systems by receiving continuous training examples.  StackRec allows us to smoothly instantiate a deeper and larger model and immediately apply it in our lifelong learning system.
We refer to such a continual or lifelong learning scenario as CL.

The second scenario is that we assume one would like to train a deep model from scratch. In such a case, StackRec also makes sense in accelerating the overall training time. Similarly, we  first train a shallow model for some steps, next copy and stack layers of it into a deep model, and then perform further training until convergence. 
We denote such a training-from-scratch scenario as TS. In the third scenario, we apply the trained deep SR model to solve the user cold-start recommendation problem, strictly following~\cite{yuan2020parameter}. Here we want to investigate whether user representations trained by StackRec could be successfully transferred to other downstream tasks.
We refer to this transfer learning scenario as TF.
\par 
We summarize our main contributions as follows.

\begin{itemize}
 	\item We shed two insights to motivate this work. 
	First, we find that the state-of-the-art SR model (e.g., NextItNet) with a small modification on its network could be stacked up over 100 layers to achieve its best  accuracy. This distinguishes from
	most existing works that apply usually less than 10 layers for their RS models.
	We believe this work is the first to  clearly
demonstrate that very deep networks are also necessary for the recommendation task, and
the network of SR models is actually allowed to be increased as deep as a computer vision model.
	Second, we find that there exists highly similar distributions between these intermediate layers/blocks in very deep SR models.
	\item We propose StackRec  to accelerate the learning process of  deep SR models. Specifically, we design two intuitive block-wise stacking methods and instantiate them by using NextItNet \cite{yuan2019simple} for case study, and report results for NextItNet, 
	the encoder of GRec~\cite{yuan2020future} (simply call it GRec),
	SASRec~\cite{kang2018self} and SSEPT ~\cite{wu2020sse} so as to verify both the effectiveness and adaptability of StackRec.
	\item We apply and evaluate StackRec in three common recommendation scenarios, namely, CL, TS and TF.
	\item Without suffering a loss in accuracy, we obtain a significant convergence speedup using StackRec  compared to the vanilla training method in all three recommendation scenarios. 
	StackRec is intuitively  simple, easy to implement and applicable to a broad class of deep SR models.
\end{itemize}


\section{Related Work}
\subsection{Deep Sequential Recommendations}

In general, deep learning (DL) based methods can be divided into three categories: RNN-based~\cite{hidasi2015session,tan2016improved,li2017neural,quadrana2017personalizing}, CNN-based~\cite{tang2018personalized,yuan2019simple}, and pure attention-based approaches~\cite{kang2018self,sun2019bert4rec}. 
RNN is a natural choice for modeling the sequential data. Specifically,
\cite{hidasi2015session} proposed GRU4Rec, which is the first work that used  RNN for the SR tasks. Following this direction, many RNN variants have been proposed, such as, GRU4Rec+ \cite{tan2016improved} with data augmentation,
personalized SR with hierarchical RNN~\cite{quadrana2017personalizing}, context-aware SR~\cite{beutel2018latent},  ranking-oriented SR~\cite{hidasi2018recurrent}.
A main drawback of RNN-based models is that they cannot be stacked very deep. 
By contrast, CNN- and attention-based models are in general deeper and more powerful than RNN models.
For example, \cite{yuan2019simple} developed NextItNet, a very deep SR model with dilated convolutions which are specifically designed for modeling  long-range item dependence. 
Recent work in~\cite{wang2019towards,yuan2020future,sun2020generic,chen2021user} showed that NextItNet usually performed better than GRU4Rec/GRU4Rec+ under a fair comparison setting.
Meanwhile, attention-based methods such as
SASRec~\cite{kang2018self} and BERT4Rec~\cite{sun2019bert4rec} that are built directly upon the Transformer~\cite{vaswani2017attention} architecture have also shown competitive performance  in the recommendation literature.  
\subsection{Accelerating Learning via Knowledge Transfer}
Improving training efficiency of deep models without sacrificing their performance has attracted much attention in the field of general machine learning~\cite{zhang2016efficient,chen2015net2net,zhou2020go,lan2019albert,gong2019efficient,li2020shallow}. To be specific,
\cite{chen2015net2net} proposed a Net2Net technique based on function-preserving transformations (FPT), which accelerated the experimentation process by instantaneously transferring the knowledge from a previously trained network to a new deeper or wider network. 
In a similar spirit, ~\cite{cai2017efficient} improved the training efficiency of Net2Net by applying reinforcement learning agent as the meta-controller, which takes actions to grow the width and depth of deep networks.
More recently, \cite{zhou2020go}  proposed a ``go WIde, then Narrow'' (WIN) method for pursuing compact and  effective deep networks.  WIN firstly widens the deep thin network, training it until convergence, and then uses the well-trained deep wide network to warm
up (or initialize)  the original deep thin network. By further fine-tuning,  WIN is able to achieve comparable performance even using a smaller-size network. From the knowledge transfer perspective,
 StackRec is similar to WIN, with the key difference that StackRec transfers knowledge from a shallower network to a deeper one whereas WIN transfers knowledge from a wider network to a thinner one. This leads to completely different applications, i.e., StackRec targets at efficient training, whereas WIN aims at model compression.  More recently, ~\cite{li2020shallow}  proposed a similar  shallow-to-deep training model for the neural machine translation task.
However, to our best knowledge, this is the first work to use  knowledge transfer techniques to accelerate training of deep SR models.  

\section{Preliminary}
\subsection{Task Definition}
Sequential recommendation aims to predict the
next item based on an ordered history of interacted items within a
user session. 
Assume that there are $|U|$ unique users $U = \{u_{1}, u_{2}, \ldots, u_{|U|}\}$ and $|I|$ unique items $I = \{i_{1}, i_{2}, \ldots, i_{|I|}\}$. 
Given a sequence of historical user behaviors $X^u=[x^u_1, x^u_2, \ldots, x^u_t]$ for user $u\in U$ in the current session, the goal of SR is to infer the item $x^u_{t+1} \in I$ that the user $u$ will interact with at time $t+1$. 
Formally, to model the joint distribution of item sequence $X^u=[x^u_1, x^u_2, \ldots, x^u_t]$, we can factorize it as a product of conditional distributions by the chain rule~\cite{yuan2019simple}: 
\begin{equation}
\label{selfsupervise}
p(X)=\prod_{i=2}^{t} p(x_{i} | x_{1: i-1},\theta) p(x_1)
\end{equation}
where $\boldsymbol{\theta}$ denotes the set of parameters for the SR model. The above optimization is often referred to as the self-supervised learning~\cite{yuan2020one,yuan2020parameter}. 
In practice, the SR model typically makes more than one recommendation by selecting the top-N (e.g., $N=5$) items from $I$, referred to the top-N sequential recommendation. 
\subsection{Base Sequential Recommendation Model}
\label{nextitnet}
We describe StackRec by specifying NextItNet \cite{yuan2019simple} as the base SR model for case study but report important results including another three deep models in literature.
NextItNet is composed of a stack of dilated convolutional  (DC)  layers, every two of which
are wrapped  by a residual block structure. To be specific,
each input item $x_i^u$ is transformed into an embedding vector  $\mathbf{e}_i^u$, and the item sequence $X^u$ is represented by an embedding matrix $\mathbf{E}^u= [\mathbf{e}_1^u, \ldots, \mathbf{e}_t^u]$. 
Then the item embeddings $\mathbf{E}^u$ are fed into a stack of 2$L$ DC layers (or $L$ DC blocks), 
which are expected to capture both short- and long-term dependencies.  Formally, the $L$-th residual block is given below:

\begin{equation}
 \label{resnet}
\begin{aligned}
   \mathbf{H}^u_{L} &= f (\mathbf{H}^u_{L-1};\Theta_L)\\
   &= \alpha_{L-1}\mathcal{F}_{L}(\mathbf{H}^u_{L-1}) + \mathbf{H}^u_{L-1}
\end{aligned}
\end{equation} 
where $\mathbf{H}^u_{L-1}\in \mathbb{R}^{t\times k}$ and $\mathbf{H}^u_{L}\in \mathbb{R}^{t\times k}$ are  the input and output of the $L$-th residual block respectively, $\mathcal{F}_{L}(\mathbf{H}^u_{L-1})$ is
residual mapping to be learned,
 $k$ is the embedding size or channel width, and $\Theta_L$ represents all parameters in the $L$-th block. Slightly different from the original NextItNet, we employ a learnable weight $\alpha_{L-1}$ for $\mathcal{F}_{L}(\mathbf{H}^u_{L-1})$ in Eq.~\ref{resnet}, where $\alpha_{L-1}$ is initialized with zero so as to ensure an identity function at the beginning of training. Our idea here trivially satisfies dynamical isometry in~\cite{chen2021user,xiao2018dynamical}. By such a simple modification, we find that NextItNet obtains not only faster convergence but also better accuracy on all our training datasets.\footnote{Note beyond the concern of this paper, we have evaluated the effectiveness of $\alpha$ for NextItNet, GRec, SASRec and SSEPT on more than 10 large-scale SR datasets, and obtain a consistent finding --- with $\alpha$, they could achieve around 1-5\% accuracy improvement w.r.t. the popular top-N metrics. Even with shallower layers, these SR models with $\alpha$ still perform better and converge faster than the original versions. 
 Some results are also shown in Section~\ref{StackRecinCL} and  Section~\ref{trans}.
}
Unless otherwise specified, NextItNet throughout this paper refers to it with the design of Eq.~\ref{resnet}.
$\mathcal{F}_{L}(\mathbf{H}^u_{L-1})$ is formed of two DC layers, defined as:
\begin{equation}
\mathcal{F}_{L}(\mathbf{H}^u_{L-1}) =\sigma\left(\mathbf{L} \mathbf{N}_{2}\left(\mathcal{C}_{2}\left(\sigma\left(\mathbf{L} \mathbf{N}_{1}\left(\mathcal{C}_{1}(\mathbf{H}^u_{L-1})\right)\right)\right)\right)\right)
\end{equation}
where $\mathcal{C}_1$ and $\mathcal{C}_2$ are the casual convolution operations, $\mathbf{L} \mathbf{N_1}$  and $\mathbf{L} \mathbf{N_2}$ represent layer normalization functions, and $\sigma$ is ReLU activation function.

Finally, a (sampled) softmax output layer is applied to predict the probability distribution for the next item $x^u_{t+1}$: 
\begin{equation}
p(x^u_{t+1}|x^u_{1:t}) = {\rm softmax}(\mathbf{W} \mathbf{H}^u_{L} + \mathbf{b})
\end{equation}
where $\mathbf{W}$ is the projection matrix, and $\mathbf{b}$ is the bias term.

\begin{figure}
	\centering
	\includegraphics[width=1\linewidth]{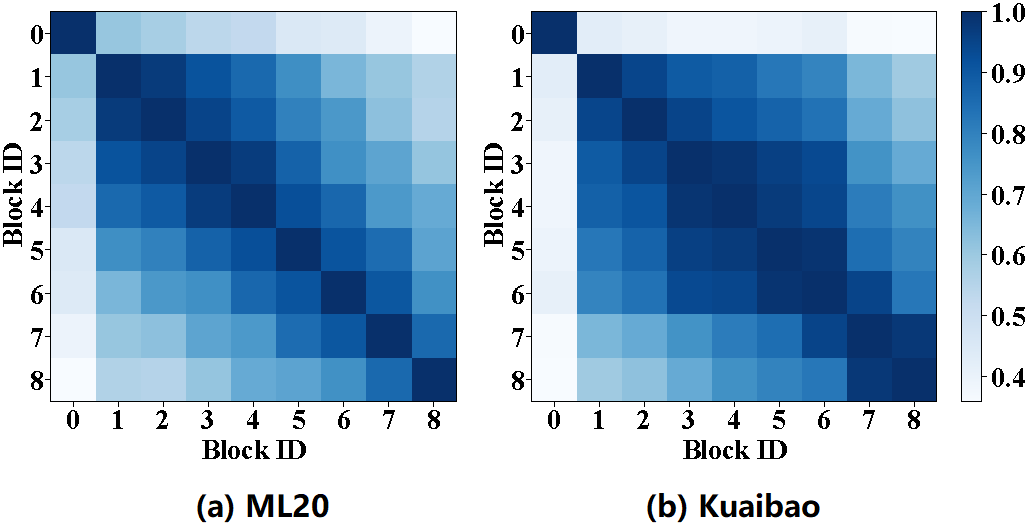}
	\caption{The cosine similarity between  feature maps of all residual blocks in NextItNet. The darker color means the higher similarity.}
	\label{similarity}
	\vspace{-0.1in}
\end{figure}


\begin{figure*}[t]
	\centering
	\subfigure[Adjacent-block stacking]{
		\includegraphics[width=0.34\linewidth]{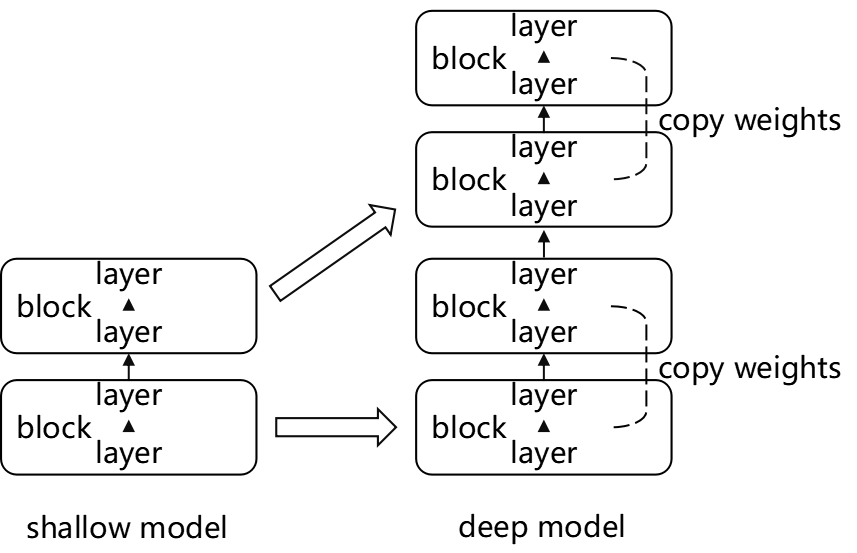}
	}\hspace{0.1\linewidth}
	\subfigure[Cross-block stacking]{
		\centering
		\includegraphics[width=0.34\linewidth]{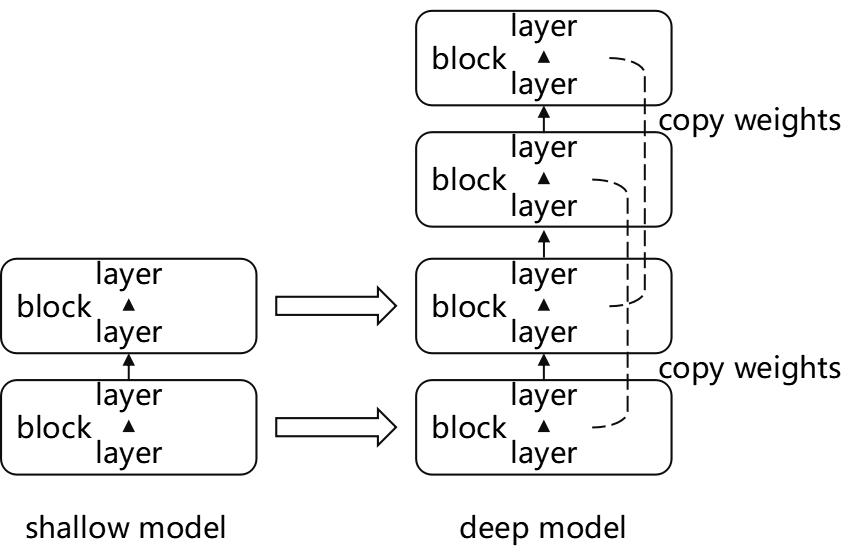}
	}
	\caption{ Two alternative block-wise stacking methods. }
	\label{stack_method}
\end{figure*}
\subsection{Motivation}

The idea of StackRec is motivated from two aspects. First, a recently proposed model compression framework called
CpRec~\cite{sun2020generic} showed that parameters of intermediate layers of a SR model could be shared across layers in many different ways. For example, top layers/blocks in CpRec are able to use the same set of parameters of the bottom layers/blocks. While such parameter sharing schemes are effective in practice, authors in~\cite{sun2020generic}
did not provide insights or explanations regarding this phenomenon.

Here, we deeply investigate reasons of the above-mentioned phenomenon by analyzing
the potential relations of different (residual) blocks of a SR model, i.e, NextItNet.
We compute cosine similarity of the output feature maps in every two successive residual blocks (or the input and output feature maps of each residual block). Specifically, we train two separate NextItNet models with 8 dilated residual blocks on the training set of ML20 and Kuaibao, respectively. After convergence, we output the feature maps of 100 randomly selected user interaction sequences from the testing set of each dataset.
We compute cosine similarity of them, i.e.,  $\mathbf{H}^u_{L-1}$ and $\mathbf{H}^u_{L}$,
and report the average of these interaction sequences on Figure \ref{similarity}. As shown, (i)
the output feature maps of every two adjacent blocks are highly similar, with the similarity
value over 90\% starting from the 2nd residual block.  This potential implies that the functionalities of these blocks  are similar to some extent.
(ii) The  similarity with the first residual block is much lower for all other blocks. We believe this is reasonable since the embedding matrix $\mathbf{E}^u$ could be largely changed after the first residual block. However, starting from the 2nd residual block, such changes affected by residual networks become smaller and smaller.  

Both our observation and the results of CpRec suggest that learned knowledge of the intermediate layers could be shared in some way due to the high similarity of them. In fact, similar observations are also made in some other domains~\cite{li2020shallow,gong2019efficient}.
This motivates us to think whether we can first train a shallower model with $L$ blocks, and then copy parameters of it and stack them into a deeper model with $2L$ blocks. This may help us substantially reduce the convergence time with good parameter initialization for these newly added top layers.
Moreover, if such mechanism works well, then we could possibly perform stacking iteratively so as to obtain a very deep SR model. In the next section, we will introduce  StackRec and its application in three common recommendation scenarios.

\section{Methodology}
\subsection{StackRec Architecture}



The core idea of StackRec is based on the progressive stacking strategy of a `shallow' SR model. 
In general, the training procedures of it consist of the following stages: (i) pre-training a shallow SR model; (ii) stacking layers of the shallow model into a double-depth model; (iii) fine-tuning this deep model; (iv) treating the deep model as a new shallower one and
 repeating (i) - (iii) until it meets the requirement.

In this paper, we propose two block-wise stacking methods for (ii), namely the adjacent-block stacking and cross-block stacking.  Suppose we have a $L$-block pre-trained model (i.e., the  shallow model). Then, we can construct a 2$L$-block model by copying the parameters of this shallow model. We perform the adjacent-block stacking in the following way: for each $i$ $<=$ $L$, the $(2i-1)$-th block and  $(2i)$-th block of the new deep model have the same parameters of the  $i$-th block of this shallow model.
Alternatively, we can perform the cross-block stacking in a slightly different way: 
for each $i$ $<=$ $L$, the $i$-th block and  $(i+L)$-th block of the same deep model have the same parameters of the  $i$-th block of the shallow model. It is also worth noting that in our StackRec framework, parameters of the embedding layer and the softmax layer of the shallow SR model should always be reused by the deep model.
For better understanding, we visualize the two stacking methods in Figure \ref{stack_method} (a) and (b), where $L$  is assumed to equal to 2. Denote $g(\cdot )$ as the function of this $L$-block network, we have
\begin{equation}
 \label{gH}
\begin{aligned}
      \mathbf{g}(\mathbf{H}^u_{0}) &= f(f(f (\mathbf{H}^u_{0};\Theta_0);\Theta_1);\ldots;\Theta_L)
\end{aligned}
\end{equation} 
Figure \ref{stack_method} (a) describes the schematic of the adjacent-block stacking method. By this way, we
make a copy of parameters in each block, first stacking the original block and the copied block, and then stacking all four blocks in the original order. Formally, the new function $G(\cdot)$ after stacking is represented as 
\begin{equation}
 \label{Ga}
\begin{aligned}
   \mathbf{G}(\mathbf{H}^u_{0}) &=f(f(f(f(f(f (\mathbf{H}^u_{0};\Theta_0);\Theta_0);\Theta_1);\Theta_1);\ldots;\Theta_L);\Theta_L)
\end{aligned}
\end{equation} 
Figure \ref{stack_method} (b) describes the schematic of the cross-block stacking method. In this way,  we can make a copy of all blocks and parameters, and then stack these two identical networks. Formally, $G(\cdot )$ in (b) is represented as 
\begin{equation}
 \label{Gb}
\begin{aligned}
   \mathbf{G}(\mathbf{H}^u_{0}) &= f(f(f(f(f(f (\mathbf{H}^u_{0};\Theta_0);\Theta_1);\ldots;\Theta_L);\Theta_0);\Theta_1);\ldots;\Theta_L)
\end{aligned}
\end{equation} 
Our adjacent/cross-block stacking methods perfectly echo the block-wise parameter sharing mechanisms in CpRec~\cite{sun2020generic}.
By stacking, the parameters copied from the trained shallow model could be a good warm-start for the constructed deep model. By fine-tuning, StackRec could
 quickly catch up the optimal performance of the same deep model which is trained from scratch in the standard way.
By recursion, StackRec that iteratively performs the above stacking technique  is supposed to obtain a very deep model faster.




\begin{figure}
	\centering
	\includegraphics[width=1\linewidth]{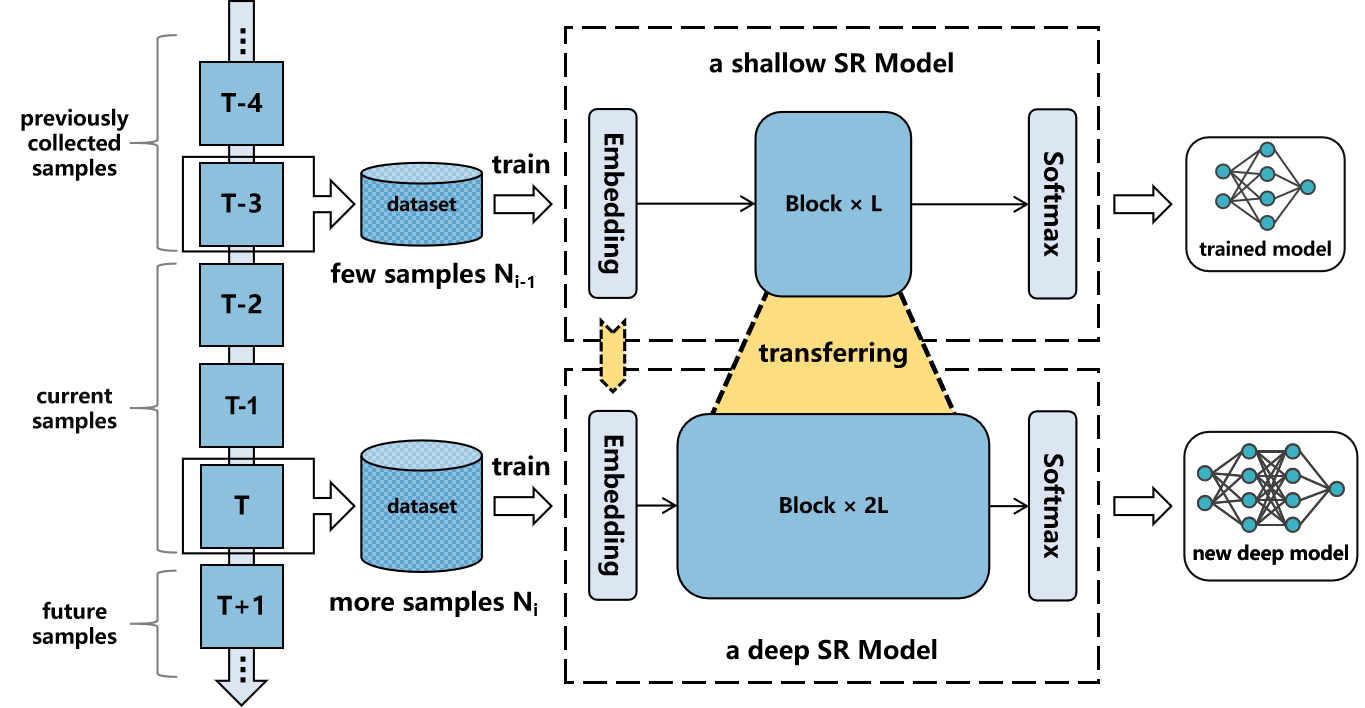}
	\caption{The architecture of StackRec in the CL scenario.}
	\label{overview}
\end{figure}
\subsection{Efficient Training in the CL Scenario}


In practice, a recommender system usually has to experience the cold-start period when it is just deployed online. In such a case, most items have very few user interactions, and as a result, the training data tends to be  highly sparse.
Empirically,  it is often sufficient to train a shallow RS model under such a sparse setting, as we illustrated in Figure~\ref{increaselayers} (a).
However, with a huge volume of new interaction data comes in, the top-performing shallow model is not expressive any more. In other words, we have to deepen the shallow model with more hidden layers so that it has higher model capacity and achieves better recommendation accuracy.
According to existing literature~\cite{zhang2020retrain}, there are two widely used solutions to handle such a CL scenario: (i) construct a deeper model and train it from scratch, and (ii) construct a deeper model and train it by utilizing previously learned parameters of the shallow model to initialize its bottom layers.  In our experiment section (see Section~\ref{two_stacking_methods}), we would compare StackRec with the two baselines.

We describe StackRec in the CL scenario and show the overall architecture in Figure~\ref{overview}. 
Assume $N_{i} \in \{ N_{0}, \ N_{1}, \ \ldots ,\ N_{k} \}$ is the training samples, and $k$ is the 
time quantum of the system when collecting the corresponding training samples.
The recommender system starts from the cold-start time with $N_0$ and
a randomly initialized shallow model NextItNet with $L$ blocks, denoted as $M_{0}$. We first train $M_{0}$ until convergence. Then, $M_{0}$ could serve the system for a period of time. Once the system has accumulated more training data $N_1$ ($N_1$ contains $N_0$), $M_{0}$ is no longer effective as before due to its shallower structure. At this stage, we expect to obtain a deeper and more expressive model $M_{1}$ with 2$L$ blocks. To this end, we perform stacking operation and double the depth of NextItNet.
After stacking, the deep NextItNet often requires fine-tuning on $ N_1$ for a while, due to a sudden change of the network connectivity. The above stacking and fine-tuning operations could be executed several times in a practical RS since a large amount of new training data is produced everyday. Algorithm~\ref{cs} illustrates such progressive stacking process in a practical system. In fact, following this way, StackRec could be regarded as a lifelong learning model which smoothly instantiates the deep model for immediate usage.

\begin{algorithm}[t]
	\caption{Progressive stacking in the CL scenario}
	\label{cs}
	\small
	\LinesNumbered
	\KwIn{the initial number of blocks $L$, stacking times $k$, training samples $\{ N_{0}, N_{1}, \ldots , N_{k} \}$ }
	\KwOut{trained model $M_{k}$ with $2^k$$*$$L$ blocks}
	$M_{0}^{*} \leftarrow$ InitNextItNet ($L$) \{ Randomly initialize NextItNet. \} \\
	$M_{0} \leftarrow$ Train ($M_{0}^{*}$) \{ Train from scratch until convergence with training samples $N_{0}$. \} \\
	\For{$i \leftarrow 1$ to $k$}{
		$M_{i}^{*} \leftarrow$ Stack ($M_{i-1}$) \{ Double the number of blocks. \} \\
		$M_{i} \leftarrow$ Train ($M_{i}^{*}$) \{
		Train with $N_{i}$ until convergence. \} \\
	}
	return $M_{k}$
\end{algorithm}

\begin{algorithm}[t]

	\caption{Progressive stacking in the TS scenario}
	\LinesNumbered
	\label{TSScenario}
	\small
	\KwIn{the desired number of blocks $2^k$$*$$L$, the initial number of blocks $L$, stacking times $k$, training samples $N$, the number of training iterations $\{ Q_{0}, Q_{1}, \ldots , Q_{k} \}$ }
	\KwOut{trained model $M_{k}$ with $2^k$$*$$L$ blocks}
	$M_{0}^{*} \leftarrow$ InitNextItNet ($L$) \{ Randomly initialize NextItNet. \} \\
	$M_{0} \leftarrow$ Train ($M_{0}^{*}$) \{ Train from scratch for $Q_{0}$ iterations with  training samples $N$. \} \\
	\For{$i \leftarrow 1$ to $k$}{
		$M_{i}^{*} \leftarrow$ Stack ($M_{i-1}$) \{ Double the number of blocks. \} \\
		$M_{i} \leftarrow$ Train ($M_{i}^{*}$) \{
		Train with $N$ for $Q_{i}$ iterations. \} \\
	}
	return $M_{k}$
\end{algorithm}

\subsection{Efficient Training in the TS Scenario}
In practice,  we may need to train a deep model directly rather than always as in the CL scenario  with a pre-trained shallow model available. In this case, we could still perform stacking like in the CL scenario. There are two main differences from the CL scenario: (i)  the shallow SR model is trained with all data rather than partial data; and (ii) it is not necessary and not allowed to train the shallow model to convergence.
Empirically, if we know the training iterations required for convergence, then we just need to train this shallow model with around $1/8 \sim 1/3$ training iterations. This helps StackRec further reduce the training time since training shallower models is much faster. Note that since both the shallow and  constructed deep models are trained with the same data, training the shallow model to convergence will make the deep model overfit quickly before finding its optimal parameter space.
Algorithm~\ref{TSScenario} illustrates the progressive stacking process in the TS scenario.



\subsection{Efficient Training in the TF Scenario}
Since the SR models are often trained by the self-supervised manner (i.e., Eq.~\ref{selfsupervise}), the learned representation can thus be used for serving various other tasks, such as the cold-user setting for item recommendations. \cite{yuan2020parameter} is the first recommendation work that demonstrates such transfer learning effect. Inspired by it, we are interested in exploring whether StackRec can also be used as a generic pre-trained model which works as effectively as NextItNet in the downstream task. In the TF scenario, we can train StackRec  following either the CL or  TS learning procedure. 
In this paper, we report results of StackRec following the procedure of the CL scenario.

\section{Experimental Setup}
\label{experimental_setup}

\subsection{Datasets}
\label{DataInfo}
We conduct extensive experiments on three real-world datasets with different session lengths. 
\begin{table}
	\caption{Statistic of the experimental datasets. ``M" and ``K" are short for million and kilo, ``t" denotes the maximum length of interaction sequences. For the ColdRec dataset, the left and right values of ``/" denote the source and target datasets, respectively.}
	\label{dataset}
	\begin{tabular}{p{0.95cm}|c|c|c|c}
		\hline
		Dataset & \# items & \# actions & \# sequences & $t$ \\
		\hline
		\hline
		ML20 & 24K & 27.5M & 1460924 & 20 \\
		\hline
		Kuaibao & 64K & 22.7M & 1000000 & 30 \\
		\hline
		ColdRec & 191K/21K & 82.5M/3.8M & 1649095/3798114 & 50/1$\sim$3\\
		\hline
	\end{tabular}
\end{table}

\begin{itemize}
	\item \textbf{ML20}: It was provided by MovieLens\footnote{https://grouplens.org/datasets/movielens/latest/}, which is widely adopted for both non-sequential and sequential recommendations~\cite{sun2020generic,sun2019bert4rec,kang2018self,tang2018personalized}. Following \cite{tang2018personalized, yuan2020future}, we perform basic data pre-processing by filtering out items with less than 10 users and users with less than 5 items
	to reduce the impact of cold-start problem. 
	We then generate the interaction
	sequence of a user according to the chronological order. 
	The maximum length of each user sequence is defined as $t$, where $t$ is set to 20. We split the sequences that exceed $t$ into multiple sub-sequences; while the sequences shorter than $t$ will be padded with zero in the beginning of the sequences to reach $t$, similar to \cite{yuan2019simple,sun2020generic}.
	\item \textbf{Kuaibao}: This is a  feed\footnote{https://kuaibao.qq.com/} (including news, micro-videos and ads) recommendation dataset collected by Tencent in June, 2019. Cold items have been simply trimmed. Each user has at most 30 recent interactions, and we set  $t$ to 30 with zero padding, similarly as in ML20. Kuaibao will be released with our source code.
	\item \textbf{ColdRec}: This is a publicly available dataset shared by PeterRec~\cite{yuan2020parameter}, which is used for evaluation in the TF scenario. 
	It contains a source-domain dataset and a target-domain dataset, which can be used to evaluate the domain adaptation ability of the SR model.
	The source-domain dataset is the news recommendation data collected from QQ Browser platform\footnote{https://browser.qq.com}, where each instance is a user's interaction sequence with at most 50 actions.
	The target-domain dataset is collected from Kandian\footnote{https://sdi.3g.qq.com/v/2019111020060911550}.  All users in Kandian are cold with at most 3 interactions (clicking of news, videos or advertisements). Each user in the target-domain has corresponding interaction records of the source-domain. 
\end{itemize}
Table \ref{dataset} summarizes the statistics of each dataset, where each instance unit is a user session sequence. 
Following~\cite{rendle2012bpr,yuan2019simple}, we randomly 
split all data into 80\% sessions for training and 20\% for evaluation.

\subsection{Baseline Models}
We compare StackRec to
its base model (NextItNet) with the same layer depth and hyper-parameters, but 
trained in the standard way. In addition, we also compare NextItNet with a well-known non-sequential recommender model NFM~\cite{he2017NFM}, and two shallow but very representative SR models Caser~\cite{tang2018personalized} and GRU4Rec~\cite{hidasi2015session} so as to demonstrate the power of deep models. For a fair comparison, we train GRU4Rec by Eq.~\ref{selfsupervise}, which shows comparable results with GRU4Rec+~\cite{tan2016improved} that applies the data augmentation~\cite{yuan2020future} method during training.
We provide the model descriptions in our experiments as below:

\begin{itemize}
    \item \textbf{SNextItNet-$k$}: It denotes the standard NextItNet without $\alpha$ in the residual block (see Section~\ref{nextitnet}). The suffix $-k$ marks the number of blocks. 
    \item \textbf{NextItNet-$k$}: It denotes NextItNet with  $\alpha$ in the residual block. Again, without special mention,  NextItNet refers to it with $\alpha$ throughout this paper.
    \item \textbf{CL-NextItNet-$k$}: It is a widely used baseline for the continual learning (CL) setting in real production systems. CL-NextItNet-$k$ performs learning on the well-optimized NextItNet model continually (without adding new layers) when new data arrives. 
    \item \textbf{StackE-Next-$k$}: Similar to StackRec, it deepens NextItNet by adding new layers. Unlike StackRec,
    only its embeddings are warm-started with learned parameters, while all blocks are randomly initialized. `E' in StackE-Next-$k$ means only the embedding layer is transferred.
    \item \textbf{StackR-Next-$k$}: Similar to StackRec, it deepens NextItNet by adding new layers. Unlike StackRec,
    only its embeddings and original blocks are warm-started with learned parameters, while newly added top blocks are randomly initialized.  Unlike CL-NextItNet-$k$, it is able to deepen its
     network structure when new data is coming.
    `R' in StackR-Next-$k$ means random initialization of newly added top blocks.
     \item \textbf{StackA-Next-$k$}: It denotes StackRec with the adjacent-block stacking method.
      \item \textbf{StackC-Next-$k$}: It denotes StackRec with the cross-block stacking method.
\end{itemize}
\begin{table*}[t]
	\caption{Overall performance comparison in the CL scenario. $x\%$ - $y\%$ simulates the following scenario: in the beginning,
	the recommender system has $x\%$ training data; after running a period of time, the system has  $y\%$ ($y > x$) training data, where $(y - x)\%$ denotes the new data. 
We take the third row with $x\%$ - $y\%$ = $40\%$ - $60\%$ as an example.
	StackC-Next-8 and 
	StackA-Next-8 mean that we use $60\%$ training data to optimize the 8-block StackRec.
NextItNet-4 (40\%) and NextItNet-8 (60\%) denote training NextItNet-4 and  NextItNet-8 with $40\%$
and $60\%$ training data respectively by learning from scratch.
	}
	\label{overall-performance}
    \small
	\begin{tabular}{C{2.1cm}|C{3.1cm}|C{1.1cm}C{1.1cm}C{1.1cm}C{1.1cm}|C{1.1cm}C{1.1cm}C{1.1cm}C{1.1cm}}
		\hline
		\multirow{2}{*}{Setting  ($x\%$ - $y\%$)} &\multirow{2}{*}{Model} & \multicolumn{4}{c|}{ML20} & \multicolumn{4}{c}{Kuaibao}   \\
		\cline{3-10} & & MRR@5 & HR@5 & NDCG@5 & Speedup & MRR@5 & HR@5 & NDCG@5 & Speedup  \\
		\hline
	    \multirow{6}{*}{100\% - 100\%}& MostPop  & 0.0062 & 0.0182 & 0.0092 & - & 0.0055 & 0.0175 & 0.0084 & - \\
	    
	    & NFM  & 0.0360 & 0.0665 & 0.0435 & - & 0.0301 & 0.0578 & 0.0369 & - \\
	    & Caser & 0.0762 & 0.1291 & 0.0892 & - & 0.0628 & 0.1069 & 0.0737& - \\
	    & GRU4Rec & 0.0786 & 0.1357 & 0.0929 & - & 0.0634 & 0.1074 & 0.0744 & - \\
	    & SNextItNet-32  & 0.0838 & 0.1431 & 0.0984 & - & 0.0667 & 0.1096 & 0.0772 & - \\
	    & NextItNet-32  & \textbf{0.0866} & \textbf{0.1473} & \textbf{0.1017} & - & \textbf{0.0694} & \textbf{0.1120} & \textbf{0.0797} & - \\
	    \hline  \hline
	    \multirow{5}{*}{40\% - 60\%} & NextItNet-4 (40\%) & 0.0703 & 0.1209 & 0.0827 & - & 0.0525 & 0.0916 & 0.0621 & - \\
	    & NextItNet-8  (60\%) & 0.0757 & 0.1288 & 0.0889 & 1.00 $\times$ & 0.0558 & 0.0974 & 0.0660 & 1.00 $\times$ \\
	    & CL-NextItNet-4 & 0.0710 & 0.1218 & 0.0835 & - & 0.0515 & 0.0913 & 0.0613 & - \\
	    & StackC-Next-8 & 0.0762 & \textbf{0.1306} & 0.0896 & \textbf{2.50 $\times$} & 0.0578 & \textbf{0.0997} & 0.0682 & \textbf{2.45 $\times$} \\
	    & StackA-Next-8  & \textbf{0.0766} & \textbf{0.1306} &  \textbf{0.0899} & \textbf{2.50 $\times$} & \textbf{0.0581} & \textbf{0.0997} & \textbf{0.0683} & \textbf{2.45 $\times$} \\
	 	\hline
	    \multirow{5}{*}{60\% - 80\%} & NextItNet-8 (60\%) & 0.0757 & 0.1288 & 0.0889 & - & 0.0558 & 0.0974 & 0.0660 & -\\
	    & NextItNet-16 (80\%)& 0.0816 & 0.1379 & 0.0955 & 1.00 $\times$ &  0.0642 & 0.1060 & 0.0746 & 1.00 $\times$ \\
	    & CL-NextItNet-8 & 0.0766 & 0.1310 & 0.0901 & - & 0.0557 & 0.0979 & 0.0661 & - \\
	    & StackC-Next-16  & 0.0819 & 0.1391 & 0.0961 & \textbf{2.05 $\times$}  & 0.0653 & 0.1082 & 0.0759 & 2.45 $\times$  \\
	    & StackA-Next-16 & \textbf{0.0822} & \textbf{0.1402} & \textbf{0.0965} & \textbf{2.05 $\times$} & \textbf{0.0658} & \textbf{0.1083} & \textbf{0.0763} & \textbf{3.06 $\times$} \\	    
		\hline
	    \multirow{5}{*}{80\% - 100\%} &  NextItNet-16 (80\%)& 0.0816 & 0.1379 & 0.0955 & - & 0.0642 & 0.1060 & 0.0746 & - \\
	    & NextItNet-32 (100\%) & 0.0866 & 0.1473 & 0.1017 & 1.00 $\times$ & 0.0694 & 0.1120 & 0.0797 & 1.00 $\times$ \\
	    & CL-NextItNet-16 & 0.0820 & 0.1400 & 0.0963 & - & 0.0638 & 0.1065 & 0.0744 & - \\
	    & StackC-Next-32  & 0.0869 & 0.1479 & 0.1020 & \textbf{3.21 $\times$}  & 0.0701 & 0.1132 & 0.0807 & 2.71 $\times$ \\
	    & StackA-Next-32  & \textbf{0.0872} & \textbf{0.1480} & \textbf{0.1022} & \textbf{3.21 $\times$} & \textbf{0.0708} & \textbf{0.1139} & \textbf{0.0814} & \textbf{3.29 $\times$} \\	    
		\hline
	\end{tabular}
	
\end{table*}
\begin{figure*}
	\centering
	\subfigure[60\% - 80\% on ML20]{
		\includegraphics[width=0.23\linewidth]{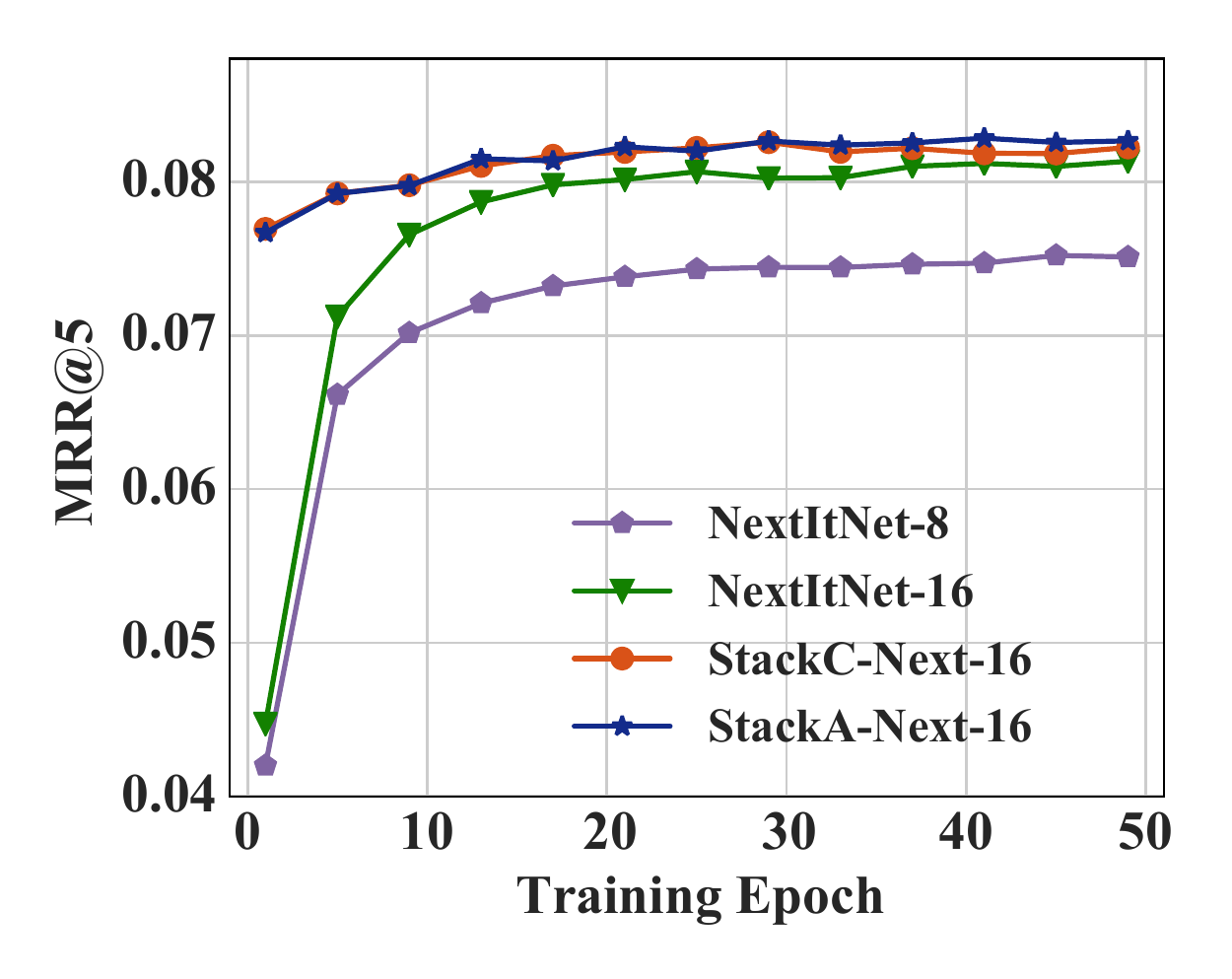}
	}
	\subfigure[80\% - 100\% on ML20]{
		\includegraphics[width=0.23\linewidth]{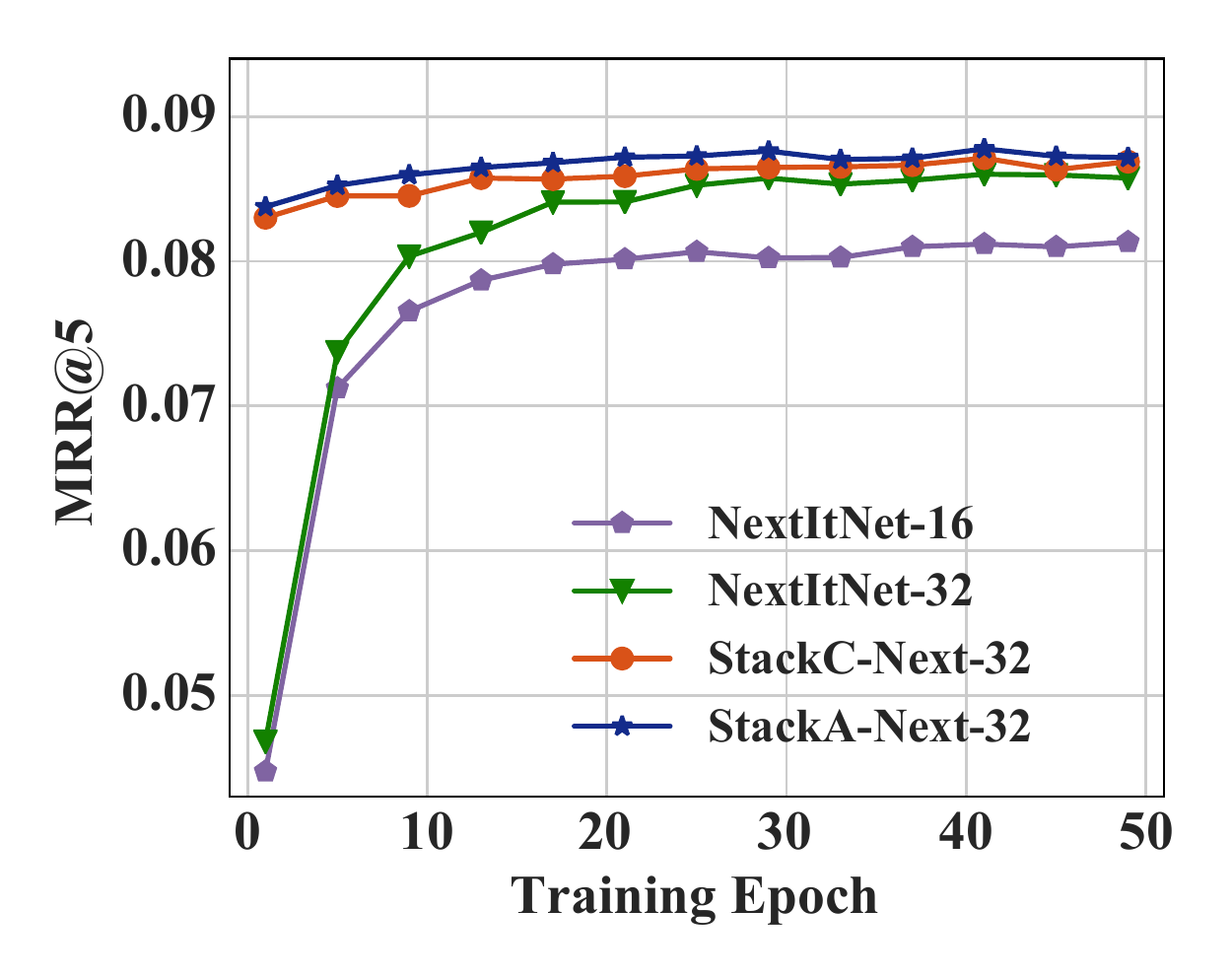}
	}
	\subfigure[60\% - 80\% on Kuaibao]{
		\includegraphics[width=0.23\linewidth]{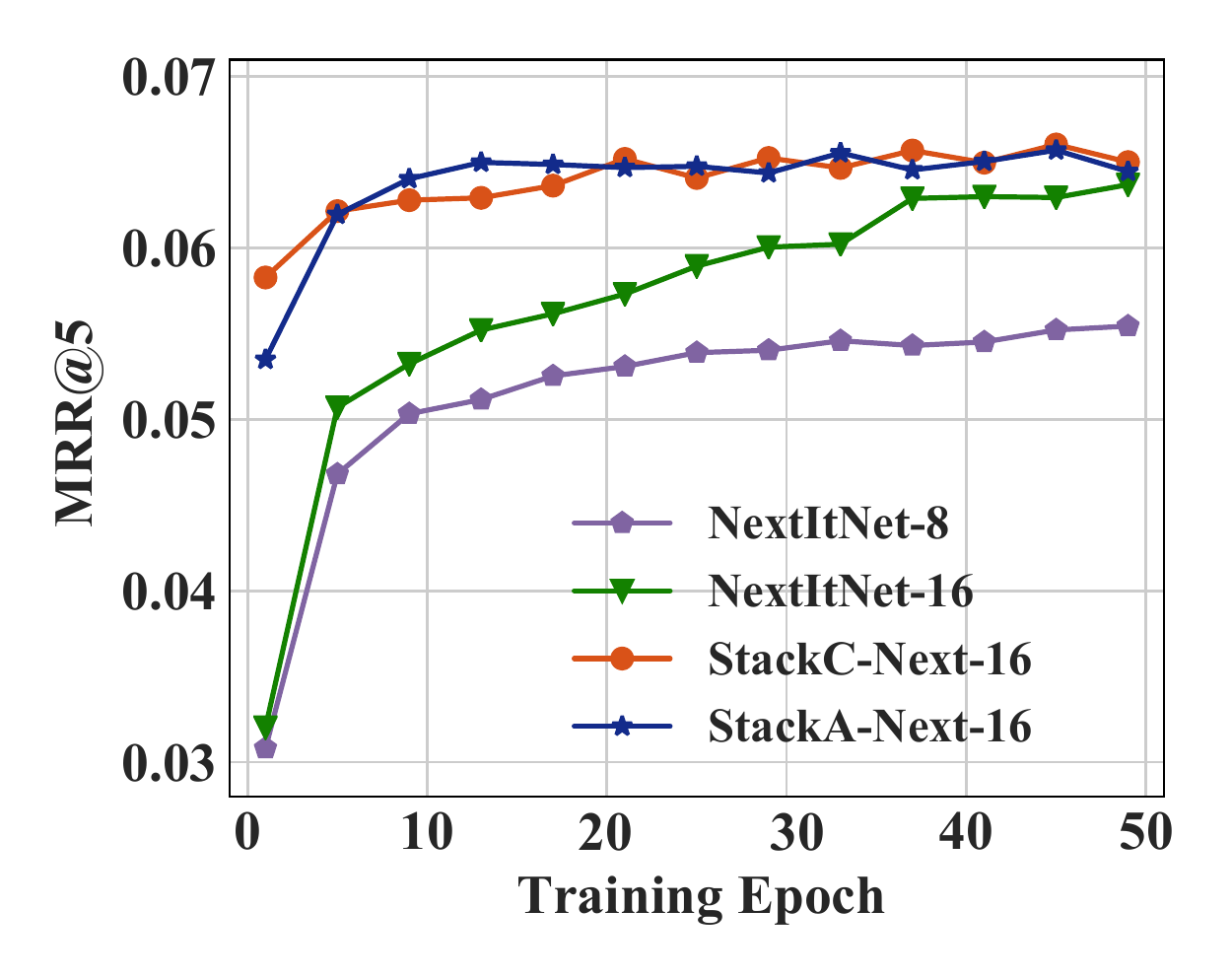}
	}
	\subfigure[80\% - 100\% on Kuaibao]{
		\includegraphics[width=0.23\linewidth]{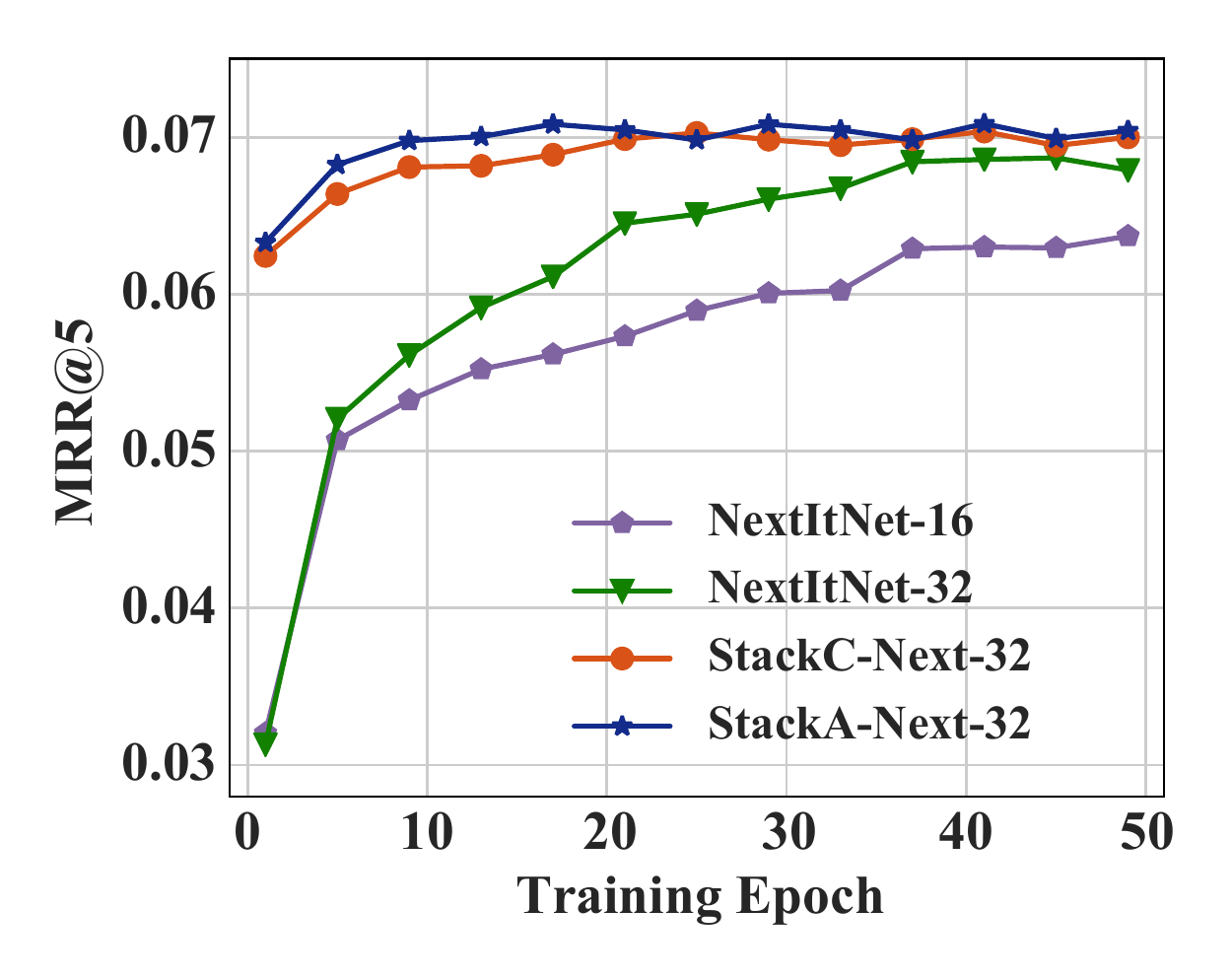}
	}
	\caption{Convergence behaviors of StackRec in the CL scenario. }
	\label{convergence}
\end{figure*}
\subsection{Implementation details}
For comparison purpose, we follow the common practice in \cite{kang2018self, yuan2019simple, rendle2014improving}  by setting the embedding dimension $d$ to 64 for all models. The dimensions of intermediate hidden layers are also set to 64. 
Although the models with different $d$ (e.g., $d=16,\ 32, \ 128$) produce different results, the performance trend keeps consistent.
The learning rate of all models is set to 0.001 on all datasets. 
The batch size is set to 256 on ML20 and Kuaibao, while it is set to 64 for
the pre-trained model and 512 for the fine-tuned model on ColdRec following~\cite{yuan2020parameter}.
Like~\cite{yuan2019simple,yuan2020future}, we use dilation factors of $\{1,2,2,4\}$ for Kuaibao and 
$\{1,2,4,8\}$ for ML20 and ColdRec. 
As for  NFM, Caser and GRU4Rec, we find that  they perform best with one hidden layer.
Other model-specific hyper-parameters for them are set based on the original paper or empirical hyper-parameter search.
All models are trained by using the cross-entropy loss and Adam \cite{kingma2014adam} optimizer.
	\vspace{-0.1in}
\subsection{Evaluation Metrics}
 Similar to \cite{kang2018self, yuan2019simple}, we only consider the prediction accuracy of the \textit{last} item in each interaction sequence in testing set for the SR task. As for the TF scenario, we follow~\cite{yuan2020parameter} by evaluating whether
 the ground-truth item in the testing set is correctly predicted.
 We adopt the most widely used evaluation metrics for top-N recommendation, including Mean Reciprocal Rank (MRR@N), Hit Ratio (HR@N), Normalized Discounted Cumulative Gain (NDCG@N) \cite{ he2020lightgcn}.
 To analyze the training efficiency, we also report the training speedup compared to the base model, denoted as Speedup. 
 \section{Experimental Results}
 In this section, we answer the following research questions:
\begin{itemize}
	\item \textbf{RQ1}: Does StackRec significantly reduce the training time of the base SR model (e.g., NextItNet) in the three scenarios? 
	If so, does it perform comparably to the base model in terms of recommendation accuracy?
	\item \textbf{RQ2}: How does StackRec perform with other intuitive or commonly used knowledge transfer baselines? 
	\item \textbf{RQ3}: Is StackRec a generic framework that works well for other typical SR models, such as GRec, SASRec and SSEPT? Does StackRec work well without $\alpha$ in the residual block?
\end{itemize}
\subsection{Overall Performance by Stacking (RQ1)}


\subsubsection{StackRec in the CL scenario.}
\label{StackRecinCL}
We report results in Table~\ref{overall-performance} and make the following observations. 
First, SNextItNet with 32 blocks (64 layers) substantially outperforms these shallow models, including NFM, Caser and GRU4Rec.
The results indicate that the deep model NextItNet/SNextItNet are more expressive when modeling user interaction sequences. Similar conclusion 
has already been made in Figure~\ref{increaselayers} (b). In addition, NextItNet largely exceeds SNextItNet, showing the effects of $\alpha$ in optimizing deep SR models.

Regarding the performance in the CL scenario, we observe that NextItNet-8 (60\%) performs better than NextItNet-4 (40\%) on both datasets, which indicates that a deeper network benefits  a lot  when more training data is available. Similar findings are also shown in other settings, i.e., 60\%-80\% and  80\%-100\%.
Nevertheless, training a deeper model from scratch requires much more computational costs and training time than a shallower model. By contrast, StackRec, including StackC-Next-8 and StackA-Next-8, achieves comparable or better results
compared to NextItNet-8 (60\%) and obtains around 2.5$\times$ speedup  in terms of training time. The results suggest that fine-tuning a deep SR model with a good warm-start substantially accelerates the training speed while maintaining a bit better recommendation accuracy. To show the convergence, we report results in Figure~\ref{convergence}. Clearly, StackRec converges much faster and a bit better than NextItNet with the same number of residual blocks. 

\subsubsection{StackRec in the TS scenario.}
\begin{figure}
	\centering
	\subfigure[training  on ML20]{
		\includegraphics[width=0.47\linewidth]{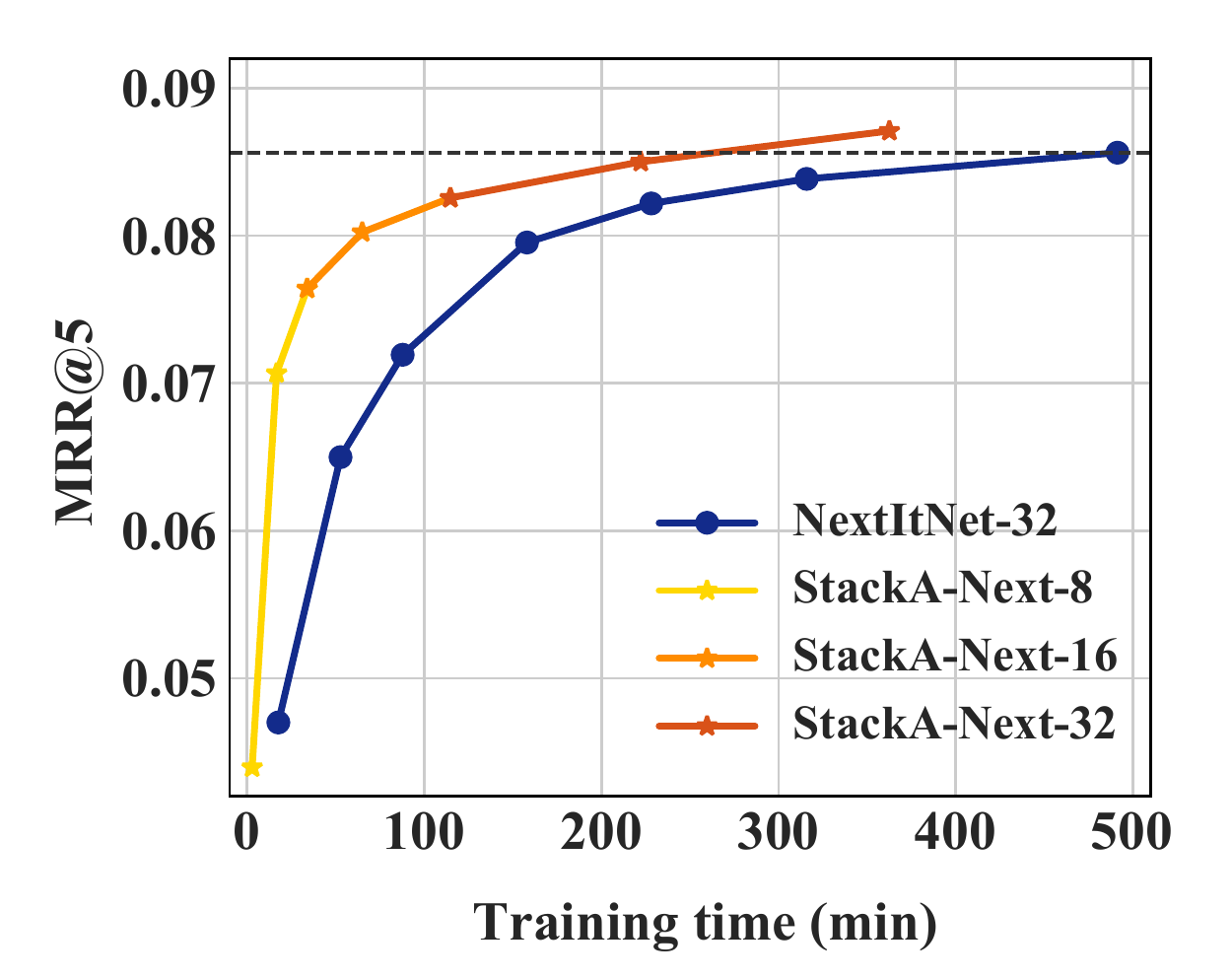}
	}
	\subfigure[training on Kuaibao]{
		\centering
		\includegraphics[width=0.47\linewidth]{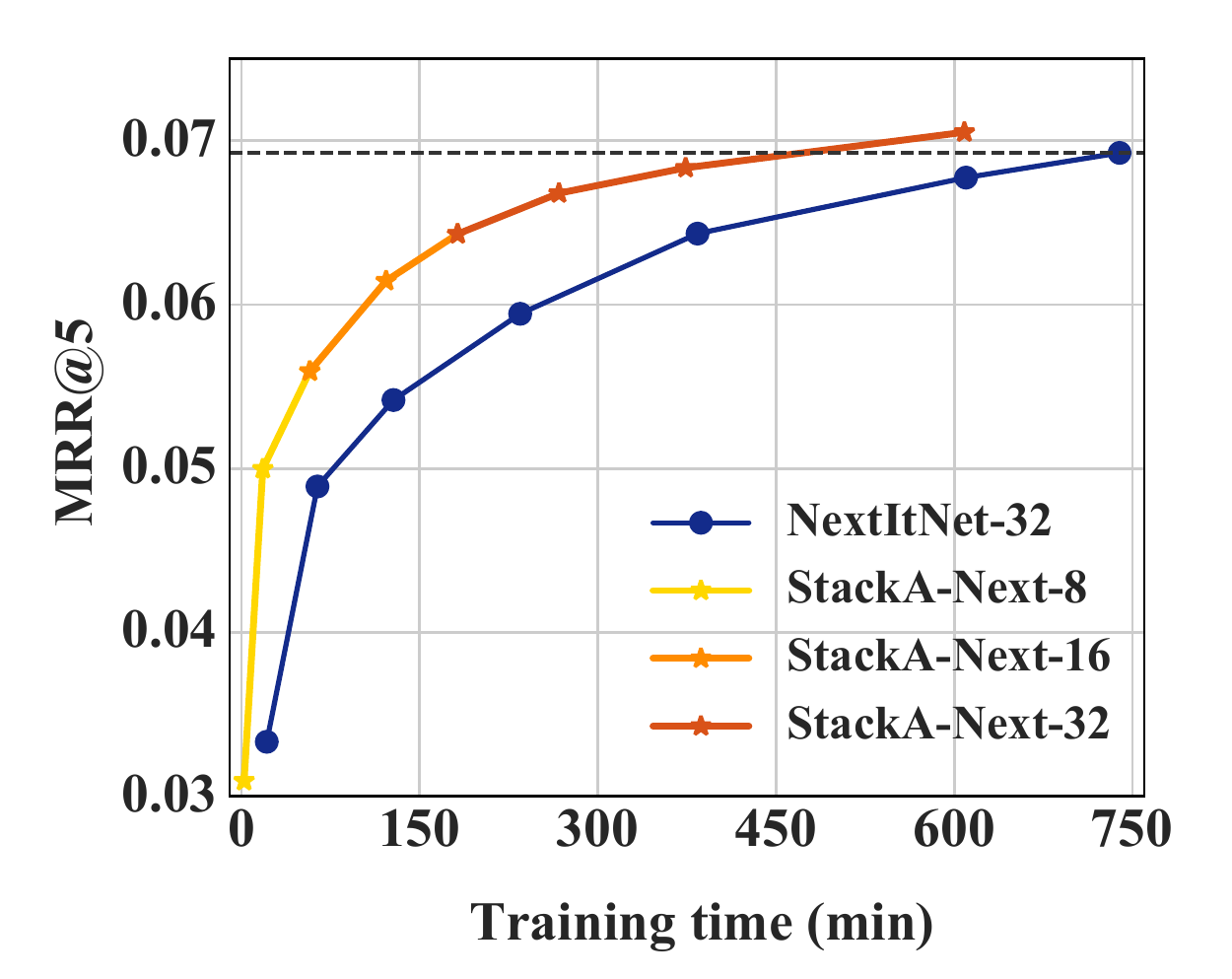}
	}
	\caption{Training curve of StackRec and the baseline in the TS Scenario. The x-axis is the wall time of training.  }
	\label{retrain_scenario}
\end{figure}

In Figure~\ref{retrain_scenario}, we show the performance changes of NextItNet and StackRec (StackA-Next-32) with training time in the TS scenario. Specifically, to obtain a 32-block StackRec, we
first train an 8-block NextItNet (training time is represented by yellow color), then deepen it into a 16-block NextItNet  by adjacent-block stacking, and perform fine-tuning (training time is represented by orange color); after that, we further deepen this 16-block NextItNet into a 32-block NextItNet and perform fine-tuning (training time is represented by red color) until convergence. As we can see,
the overall training time of StackRec is about $40\%$ shorter (280 minutes vs. 490 minutes) on ML20 and about $35\%$ shorter (480 minutes vs. 740 minutes) on Kuaibao. The main speedup is 
obtained because (i) training a shallower model takes much less time; (ii) warm-starting parameters transferred from the shallow model is of great helpful for the convergence of a deep model.

\subsubsection{StackRec in the TF scenario.}
\begin{table}
	\caption{Results of StackRec in the TF scenario. ${\rm \text{MRR}_{t}}$ and ${\rm \text{HR}_{t}}$ represent the accuracy on the target domain with MRR@5 and HR@5, respectively, whereas ${\rm \text{MRR}_{s}}$, ${\rm \text{HR}_{s}}$ and speedup are evaluated on the source domain. We only show results of setting 80\% - 100\% (of the CL scenario) for clarity. }
	\label{transfer-learning}
	\small
	\begin{tabular}{c|C{0.85cm}C{0.85cm}C{0.85cm}C{0.85cm}C{1cm}}
		 Model &  ${\rm MRR_{s}}$ & ${\rm HR_{s}}$ & ${\rm MRR_{t}}$ & ${\rm HR_{t}}$ & Speedup  \\
		\midrule
		NextItNet-8 & 0.0212 & 0.0419 & 0.2014 & 0.3497 & - \\
        NextItNet-16 & 0.0218 & 0.0430 & 0.2048 & 0.3548 &  1.00 $\times$ \\
	    StackC-Next-16 & \textbf{0.0220}  &  0.0436 & 0.2051 & 0.3570 & 3.09 $\times$ \\
	    StackA-Next-16 & \textbf{0.0220}  & \textbf{0.0437}& \textbf{0.2056} & \textbf{0.3582} & \textbf{3.58 $\times$}  \\
	\end{tabular}
	\vspace{-0.1in}
\end{table}
As mentioned before, deep learning based SR models can not only recommend items in the same system, but also work as a pre-trained network that can be adapted to solve recommendation problems from a different system.  To examine the transfer learning ability of StackRec, we first 
perform pretraining on the source-domain dataset of ColdRec using StackRec. Then we simply add a new softmax layer on the final hidden layer of StackRec, and fine-tune all parameters on the target-domain dataset by using the pre-trained weights as warm start. Our transfer learning framework and evaluation strictly follow~\cite{yuan2020parameter}. We report the results in Table \ref{transfer-learning}. 
\par
As shown, StackRec obtains around 3$\times$ speedup on ColdRec and yields comparable pre-training and fine-tuning performance compared with NextItNet-16. In practice, pre-training a large model usually requires  a large-scale dataset and longer training time. Therefore, the training efficiency of pre-trained networks becomes a critical issue for TF since the requirement of extremely high-performance hardware is a barrier to its practical application. Our StackRec is helpful for such a TF scenario.




\subsection{Ablation Studies (RQ2)}
\label{rq2}

\subsubsection{Comparison with two other stacking methods.} 
\label{two_stacking_methods}
In the CL scenario, one may come up with another two intuitive baselines: (1) deepening
the shallow network with new layers, but randomly initializing
these newly added layers (i.e., StackR-Next); (2) only transferring the embedding layer of the shallow network to the deep model (i.e., StackE-Next). The performance of StackR-Next and StackE-Next helps us to identify whether the proposed stacking method is really helpful and necessary. We report the comparison in
Table~\ref{different_stacking_methods}. As shown, StackR-Next and StackE-Next  are indeed able to accelerate the training convergence speed compared to  NextItNet-16. But they also yield worse recommendation accuracy, especially on Kuaibao. In particular, StackR-Next performs even worse than StackE-Next.
We believe that this is explainable since though warm-starting a deep network by parameters of its shallower version  reduces its convergence time, it might get trapped in some local optimum easily, and lead to worse accuracy. 
By contrast, StackRec performs much better and converges faster using our proposed stacking methods. Thereby, we conclude that warm-starting all layers like StackRec results in better generalization and accuracy.

\subsubsection{Comparison by stacking different number of blocks.} 
In previous experiments, we perform stacking by  doubling the number of blocks.
 In practice, 
 we may not always need such a deep network. Here we show that the number of blocks to be stacked can be any integer value.
 As shown in Table \ref{number_of_blocks},  NextItNet-64 only achieves less than 1\% accuracy gain than  NextItNet-48. In such a case, NextItNet-48 may meet the performance requirement of many applications. That is, we just need to stack 16  blocks rather than 32 blocks.  As shown, StackRec  by stacking 16 blocks performs as well as NextItNet-48 and obtains around 2.4$\times$ speedup.
 Therefore, we conclude that StackRec is flexible so that we can choose any number of blocks to stack based on actual demand.

\begin{table}
	\caption{Results of other stacking methods. To save space,
	we  only show results of setting 60\% - 80\% in the CL scenario.}
	\label{different_stacking_methods}
	\small
	\begin{tabular}{C{2.2cm}|C{1.1cm}C{1.1cm}|C{1.1cm}C{1.1cm}}
    \multicolumn{1}{c}{} & \multicolumn{2}{c}{ML20} & \multicolumn{2}{c}{Kuaibao} \\
     Model & MRR@5 & Speedup & MRR@5 & Speedup \\
    \midrule
    NextItNet-16 &  0.0816  & 1.00 $\times$ & 0.0642 & 1.00 $\times$ \\
    StackE-Next-16 & 0.0815 & 1.20 $\times$ & 0.0640 & 1.15 $\times$ \\
    StackR-Next-16 & 0.0807 & 1.55 $\times$ & 0.0613 & 2.35 $\times$ \\
	StackC-Next-16 & 0.0819 & \textbf{2.05 $\times$} & 0.0653 & 2.45 $\times$ \\
	StackA-Next-16 & \textbf{0.0822} & \textbf{2.05 $\times$} & \textbf{0.0658} & \textbf{3.06 $\times$} \\
	\end{tabular}
\end{table}
\begin{table}
	\caption{Results of stacking different number of blocks on Kuaibao. We use setting 80\% - 100\% in the CL scenario here.}
	\label{number_of_blocks}
	\small
	\begin{tabular}{C{2.2cm}|C{1.1cm}C{1.1cm}C{1.1cm}C{1.1cm}}
		\multicolumn{1}{c|}{Model} &  MRR@5 & HR@5 & NDCG@5 & Speedup  \\
		\midrule
		NextItNet-32 & 0.0669 & 0.1093 & 0.0773 & - \\
		NextItNet-48 & 0.0703 & 0.1133 & 0.0809 & 1.00 $\times$\\
		NextItNet-64 & 0.0709 & 0.1140 & 0.0815 & 1.00 $\times$ \\
		StackC-Next-48 & 0.0719 & 0.1146 & 0.0824 & \textbf{2.41 $\times$} \\
		StackA-Next-48 & \textbf{0.0725} & \textbf{0.1151} & \textbf{0.0830} & \textbf{2.41 $\times$} \\
	\end{tabular}
	\vspace{-0.1in}
\end{table}
\subsection{Adaptability Experiment (RQ3)}
\label{trans}
Here we investigate StackRec by using SASRec, SSEPT
\footnote{Note that the results of SSEPT are not  comparable with other models since the original paper~\cite{wu2020sse} implemented it by concating context features with item embeddings, which  results in model size with around two times larger than other models.} and the encoder of GRec~\cite{yuan2020future}, denoted as GRec for short, and report results in Table \ref{adapt}. Like NextItNet, we also add $\alpha$ in 
their residual blocks, so that these models can be stacked very deep and perform further better. 
 Again, we emphasize that all conclusions made for StackRec hold on their original versions, i.e., without $\alpha$, as reported in Table~\ref{without_alpha}.
\par
As shown, very similar conclusions can be made as in Section \ref{StackRecinCL}.  Specifically, StackRec can significantly improve the training speed for three models. The average speedup compared to their non-stack versions is around $2\times$. Also, StackRec performs comparably to SASRec, SSEPT and GRec with the same layer depth in terms of all metrics. That is, StackRec is not a specialized framework that only works for NextItNet. Instead, it can be simply applied to a broad class of deep SR models, yielding significant training acceleration.


\begin{table}
	\caption{Results of StackRec with or without $\alpha$ on ML20. We use setting 80\% - 100\% in the CL scenario here.}
	\label{without_alpha}
	\small
	\begin{tabular}{C{2.2cm}|C{1.1cm}C{1.1cm}|C{1.1cm}C{1.1cm}}
    \multicolumn{1}{c}{} & \multicolumn{2}{c}{Without $\alpha$} & \multicolumn{2}{c}{With $\alpha$} \\
     Model & MRR@5 & Speedup & MRR@5 & Speedup \\
    \midrule
    NextItNet-8 & 0.0785 & - & 0.0797 & - \\
    NextItNet-16 &  0.0832  & 1.00 $\times$ & 0.0847  & 1.00 $\times$ \\
	StackC-Next-16 & 0.0836 & 2.21 $\times$ & 0.0850 & 2.19 $\times$ \\
	StackA-Next-16 & \textbf{0.0840} & \textbf{2.46 $\times$} & \textbf{0.0852 } & \textbf{ 2.27 $\times$} \\
	\end{tabular}
\end{table}

\begin{table}
	\caption{Stacking with GRec, SASRec and SSEPT on ML20. We use setting 80\% - 100\% in the CL scenario here.}
	\label{adapt}
	\small
	\begin{tabular}{C{2.2cm}|C{1.1cm}C{1.1cm}C{1.1cm}C{1.1cm}}
		\multicolumn{1}{c|}{Model} &  MRR@5 & HR@5 & NDCG@5 & Speedup  \\
		\midrule
		GRec-16 & 0.0819 & 0.1392 & 0.0960 & - \\
		GRec-32 & 0.0867 & 0.1483 & 0.1019 & 1.00 $\times$\\
		StackC-GRec-32 & 0.0875 & 0.1488 & 0.1027  & 2.33 $\times$ \\
		StackA-GRec-32 & \textbf{0.0878} & \textbf{0.1492} & \textbf{0.1029} & \textbf{2.47 $\times$} \\
		\midrule
		SASRec-18 & 0.0894 & 0.1509 & 0.1042 & - \\
		SASRec-24 & 0.0925 & 0.1565 & 0.1083 & 1.00 $\times$ \\
		StackC-SASRec-24 & \textbf{0.0927}	& 0.1566 & 0.1085 & 2.06 $\times$ \\
		StackA-SASRec-24 & \textbf{0.0927} & \textbf{0.1569} & \textbf{0.1086} & \textbf{2.28 $\times$} \\
		\midrule
		SSEPT-18 & 0.1010 & 0.1689 & 0.1178 & - \\
		SSEPT-24 & 0.1043 & 0.1740 & 0.1215 & 1.00 $\times$ \\
		StackC-SSEPT-24 & 0.1066 & 0.1773 & 0.1241 & 2.16 $\times$ \\
		StackA-SSEPT-24 & \textbf{0.1068} & \textbf{0.1776} & \textbf{0.1243} & \textbf{2.48 $\times$} \\
	\end{tabular}
		\vspace{-0.1in}
\end{table}

\section{Conclusions}
We have thoroughly investigated the impact of network depth for deep sequential recommendation (SR) models,
and surprisingly found that stacking super deep layers \textit{could} be beneficial in practice, which largely agrees with the recent research advancement in other fields, e.g., computer vision.
We proposed StackRec, a very simple, yet effective and useful  framework to speed up the training process of deep SR models. Specifically, we introduced two progressive stacking techniques motivated by the basic fact that intermediate layers/blocks in deep SR models  have highly similar functions. Moreover, we studied StackRec in three common recommendation scenarios. Through extensive experiments on  real-world datasets, we showed that StackRec could considerably accelerate the training of deep SR models without performance drop.

\begin{acks}
This work was supported by National Natural Science Foundation of China (No. 62072186, 61876208 and 61906185), Guangdong Basic and Applied Basic Research Foundation (No. 2019B1515130001) and Key-Area Research and Development Program of Guangdong Province (No. 2018B010108002). Min Yang was partially supported by Youth Innovation Promotion Association of CAS China (No. 2020357), Shenzhen Science and Technology Innovation Program (No. KQTD20190929172835662). 
\end{acks}   

\bibliographystyle{ACM-Reference-Format}
\bibliography{sample-base}

\end{document}